\documentclass[namedreferences]{SolarPhysics}
\usepackage[optionalrh]{spr-sola-addons} 
\usepackage{graphicx}        
\usepackage{natbib}         
\usepackage{color}           
\usepackage{url}             

\newcommand{\aap}{    {\it Astron. Astrophys.}}

\newcommand{\apj}{    {\it Astrophys. J.}}

\newcommand{\jgr}{    {\it J. Geophys. Res.}}

\newcommand{\solphys}{{\it Solar Phys.}}
\newcommand{\sovast}{ {\it Sov. Astronom.}}

\newcommand{\araa}{    {\it Annu. Rev. Astron. Astrophys.}}

\begin{document}

\begin{article}

\begin{opening}

\title{Probing Twisted Magnetic Field Using Microwave Observations in an M Class Solar Flare on 11 February, 2014}

\author{I.N.~\surname{Sharykin}$^{1,2}$\sep
        A.A.~\surname{Kuznetsov}$^{2}$\sep
        I.I.~\surname{Myshyakov}$^{2}$\sep
       }
\runningauthor{Sharykin~I.N. et al.}
\runningtitle{Radio Emission From a Twisted Magnetic Loop}
   \institute{$^{1}$ Space Research Institute of Russian Academy of Sciences (IKI), Moscow, Russia
                     email: \url{ivan.sharykin@phystech.edu}\\
					$^{2}$ Institute of Solar-Terrestrial Physics, Irkutsk 664033, Russia
             }

\begin{abstract}
This work demonstrates the possibility of magnetic field topology investigations using microwave polarimetric observations. We study a solar flare of GOES M1.7 class that occurred on 11 February, 2014. This flare revealed a clear signature of spatial inversion of the radio emission polarization sign. We show that the observed polarization pattern can be explained by nonthermal gyrosynchrotron emission from the twisted magnetic structure. Using observations of the {\it Reuven Ramaty High Energy Solar Spectroscopic Imager}, Nobeyama Radio Observatory, Radio Solar Telescope Network, and {\it Solar Dynamics Observatory}, we have determined the parameters of nonthermal electrons and thermal plasma and identified the magnetic structure where the flare energy release occurred. To reconstruct the coronal magnetic field, we use nonlinear force-free field (NLFFF) and potential magnetic field approaches. Radio emission of nonthermal electrons is simulated by the \textsf{GX Simulator} code using the extrapolated magnetic field and the parameters of nonthermal electrons and thermal plasma inferred from the observations; the model radio maps and spectra are compared with observations. We have found that the potential magnetic field approach fails to explain the observed circular polarization pattern; on the other hand, the Stokes $V$ map is successfully explained by assuming nonthermal electrons to be distributed along the twisted magnetic structure determined by the NLFFF extrapolation approach. Thus, we show that the radio polarization maps can be used for diagnosing the topology of the flare magnetic structures where nonthermal electrons are injected.
\end{abstract}
\end{opening}

\section{Introduction}
During solar flares, electrons are accelerated to relativistic speeds from background thermal plasma; these energetic nonthermal electrons produce gyrosynchrotron (GS) radio emission ($\gtrsim$1~GHz) and hard X-ray (HXR) emission. In the standard model of an eruptive two-ribbon flare \citep{Hirayama1974,Magara1996,Tsuneta1997}, soft X-ray (SXR) emitting magnetic loops where plasma is heated by nonthermal electrons are formed due to magnetic reconnection in the cusp below the erupting plasmoid, causing a coronal mass ejection (CME). HXR observations made by the {\it Reuven Ramaty High Energy Solar Spectroscopic Imager} (RHESSI: \citealp{Lin2002}) reveal a lot of loop-like HXR emission sources \citep[e.g.]{Battaglia2005,Jiang2006,Guo2012}. Nobeyama Radioheliograph (NoRH: \citealp{Nakajima94}) observations also show us loop structures in the microwave range \citep[e.g.][]{Kupriyanova2010,Morgachev2014}. Loop-like geometry of the X-ray and microwave emission sources is a usual observational manifestation of the flare energy release process.

The classical two-dimensional model of magnetic reconnection assumes interaction of the opposite-polarity magnetic flux tubes at a null-point. At the reconnection site, plasma is heated and thermal electrons are accelerated, forming nonthermal power-law energetic spectrum. However, magnetic reconnection can occur in a magnetic configuration without null points as well. For example, twisted magnetized plasma loops can experience internal magnetic reconnection \citep{Demoulin1996,Gordovskyy2011,Pinto2016}; in such a case, accelerated particles will be directly accelerated and injected into the loop volume \citep{Gordovskyy2011,Gordovskyy2012,Gordovskyy2013,Gordovskyy2014}.

We know that orientation of the magnetic field in the flare region affects the spatial distribution of brightness and polarization of the radio emission. In particular, in the case of isotropic or pancake-like anisotropic pitch-angle distribution of nonthermal electrons, the extraordinary mode of radio waves is dominant and inversion of the circular polarization sign corresponds to inversion of the line-of-sight (LOS) component of magnetic field. The simplest topology of the magnetic field is described by the potential field approach; in the work of \citealp{Kuznetsov2011}, synthetic radio maps of potential magnetic loops have revealed that the polarization sign inversion line (PSIL) is perpendicular to the loop axis both for the central and limb positions of the loop. The work of \citealp{Sharykin2016} was devoted to modelling of the gyrosynchrotron radio emission of nonthermal electrons from twisted magnetic loops using \textsf{GX Simulator} \citep{Nita2015}; the authors considered an analytical Titov--Demoulin model \citep{Titov1999} of the twisted magnetic flux-rope. The work has shown that in such magnetic structures the inversion line of the radio emission polarization sign is inclined relative to the loop axis; polarization of the radio emission from a twisted loop on the solar limb demonstrates sign reversal \textit{along} the loop axis. More complex modelling of nonthermal GS emission from twisted magnetic configurations was presented in the work of \citealp{Gordovskyy2017}. To model the emission, the authors used the magnetic field structure resulting from numerical magnetohydrodynamic (MHD) simulation of the kink-instability of the twisted magnetic flux-rope; energetic and pitch-angle distributions of accelerated electrons were obtained from simultaneous test-particle modelling. It has been found that the flaring twisted coronal loops produce GS emission with a gradient of circular polarization \textit{across} the loop; these patterns may be visible only for a short period of time due to fast magnetic reconfiguration. We should also note that the limited spatial resolution of the existing radio telescopes does not allow us to investigate directly the polarization distribution across a flaring loop.

Thus, potentially, one can use the spatially-resolved microwave polarization patterns to diagnose the topology of the magnetic field where nonthermal electrons propagate. However, it is not clear whether the available microwave observations allow us to identify the features mentioned in the radio polarization maps. Currently, the best spatial resolution among the solar-dedicated radio instruments is achieved by NoRH and is about $5''$ at 34 GHz and $10''$ at 17 GHz; spatially resolved observations of both Stokes $I$ and $V$ components are made at 17 GHz only. For the small-scale magnetic structures, whose sizes are comparable with the NoRH beam size, the observed Stokes $V$ maps can differ significantly from real distributions of circular polarization in the emission source; that is why, e.g., the expected polarization patterns corresponding to the twisted magnetic loops at the limb cannot be detected with the NoRH. \citealp{Sharykin2016} presented the synthetic radio maps convolved with NoRH-like symmetric Gaussian beams and showed that it is still possible to detect the polarization peculiarities of the radio emission sources in the twisted magnetic loops located on the solar disk far from the limb. Ideally, to test the predictions of the works of \citealp{Sharykin2016} and \citealp{Gordovskyy2017}, one should select solar flares with the radio emission source larger than the NoRH beam.

The main scope of this work is studying the GS radio emission of nonthermal electrons in twisted magnetic loops during solar flares and testing the theoretical predictions described in the works of \citealp{Sharykin2016} and \citealp{Gordovskyy2017}. For this purpose, real radio observations of a flare are compared with the simulated radio images. The NoRH observations are used to determine the spatially resolved microwave polarization patterns at 17 GHz, while the Nobeyama Radio Polarimeters (NoRP), Radio Solar Network Telescope (RSTN), and RHESSI data are used to extract information about the energetic spectrum of nonthermal electrons. To identify the magnetic loops where the nonthermal electrons are injected, we use the EUV images from the {\it Atmospheric Imaging Assembly} (AIA: \citealp{Lemen12}) onboard the {\it Solar Dynamics Observatory} (SDO) and the images from the {\it X-ray Telescope} (XRT: \citealp{Golub07}) onboard the {\it Hinode} spacecraft to find the hottest part of the flare loop system. Information about the magnetic field in the flare region is extracted from the SDO {\it Helioseismic and Magnetic Imager} (HMI: \citealp{Scherrer2012}) vector magnetograms; these magnetograms are used for magnetic field extrapolation. Nonlinear force-free field (NLFFF) and potential magnetic field approaches are used for the coronal magnetic field reconstruction. All the obtained data together are used to create models for three-dimensional simulation of the flare GS radio emission with the \textsf{GX Simulator}. The results of the simulations obtained using the two extrapolation approaches are then compared with the radio observations.

\section{Observations}\label{Observations}

\subsection{Event Selection}
In this section we describe observations of the solar flare selected for analysis. There were several main criteria of the event selection:

\begin{enumerate}[vi.]
\renewcommand{\labelenumi}{\roman{enumi}.}%
\item Availability of the NoRH and NoRP observations of the nonthermal microwave emission. The microwave spectrum peak should be observed at the frequencies below 17 GHz to allow determination of the high-frequency slope, which reflects the nonthermal electrons spectral index.
\item Availability of the RHESSI HXR observations above 50 keV with a sufficient count rate needed for spectroscopy and imaging.
\item A clear spatial inversion of the polarization sign in the 17 GHz NoRH microwave images.
\item The presence of double HXR emission sources in the RHESSI images (they usually correspond to the footpoints of the magnetic loops where nonthermal electrons precipitate into chromosphere) to define the flare loop system.
\item Strong photospheric electric currents determined from the HMI vector magnetograms (see Section \ref{SDO}) near the flare emission sources; such currents are necessary for the formation of a twisted magnetic configuration.
\item Flare location in the central part of the solar disk ($\lesssim 300''$ from the center).
\item AIA images in the 94 and 335 \AA{} (the less sensitive) channels should not be saturated during the impulsive phase; this is necessary for proper detection of the flaring loops. Thus we consider moderately powerful solar flares: high C and low M classes.
\end{enumerate}

We have found several events satisfying the criteria. In this article we consider one example to develop the analysis technique: SOL2011-02-11 event of GOES M1.7 class, with the peak (according to GOES data) at 03:31 UT. The flare occurred in active region NOAA 11974 with the heliographic coordinates S13E07.

\begin{figure}
\centering
\includegraphics[width=8cm]{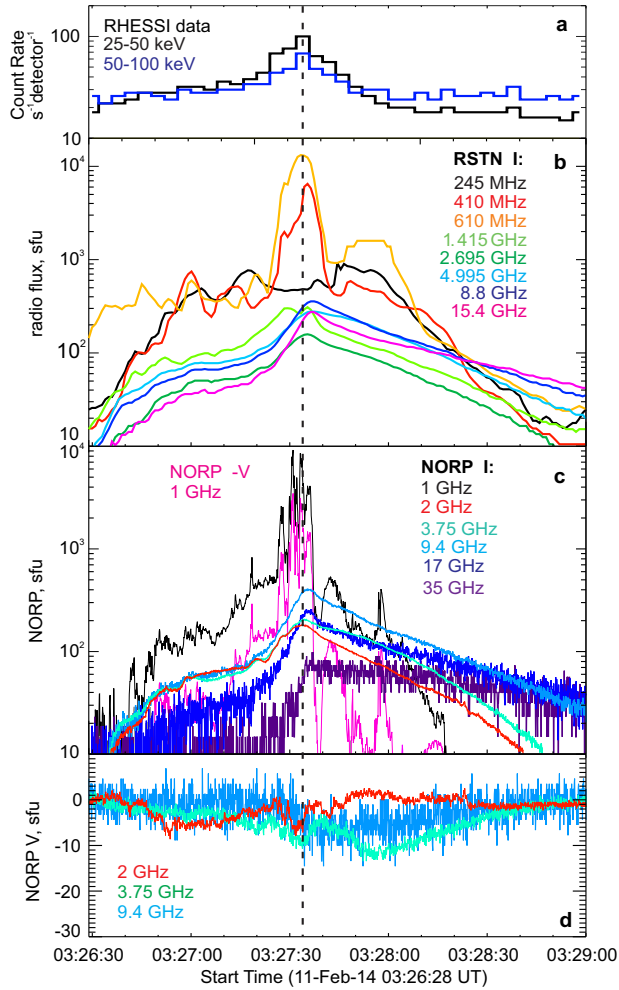}
\caption{Light curves of the M1.7 flare on 11 February, 2014 observed at HXR and radio wavelengths. a) RHESSI count rates ($25-50$ and $50-100$ keV). b) RSTN Stokes $I$ time profiles (0.245, 0.41, 0.61, 1.415, 2.695, 4.995, 8.8, and 15.4 GHz). c) NoRP Stokes $I$ time profiles (1, 2, 3.75, 9.4, 17, and 35 GHz) and unsigned Stokes $V$ time profile at 1 GHz. d) NoRP Stokes $V$ time profiles (2, 3.75, and 9.4 GHz).}
\label{TPs}
\end{figure}

\subsection{Temporal Evolution}
Figure \ref{TPs} shows the temporal evolution of the selected flare (around the impulsive phase) at different wavelengths. In particular, Figure \ref{TPs}a shows temporal profiles of the RHESSI HXR count rates in the $25-50$ and $50-100$ keV energy bands. HXR peaked at around 03:27:39~UT, and the total duration of the impulsive phase (according to the $25-50$~keV data) was about 100~seconds.

\begin{figure}
\centering
\includegraphics[width=1.0\linewidth]{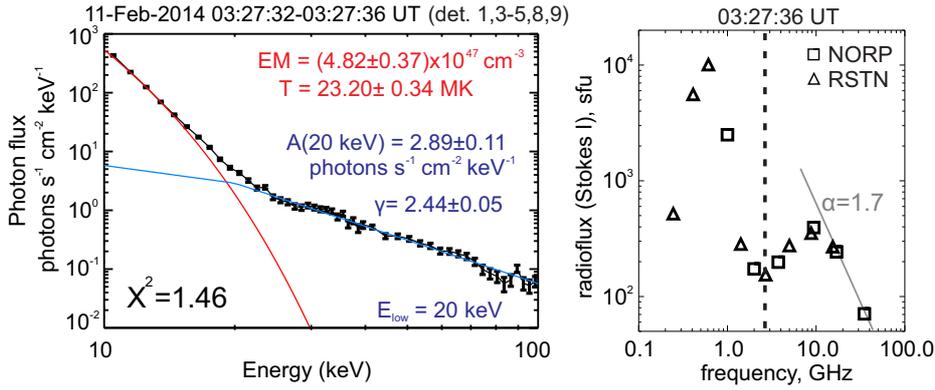}
\caption{RHESSI X-ray (left) and NoRP and RSTN radio (right) spectrum around the flare peak. In the left panel, the thermal and nonthermal (double power-law) components of the model fitting function are shown by the red and blue lines, respectively. In the right panel, the grey line connects the NoRP points at 17 and 35 GHz.}
\label{spec}
\end{figure}

Figures \ref{TPs}b\,--\,d show the NoRP and RSTN radio light curves (NoRP measures both the emission intensity and circular polarization, while RSTN records the intensity only). Radio emission peaked at the same time as the HXR; most likely, it was produced by the same nonthermal electrons. Radio observations (see also right panel of Figure \ref{spec}) reveal two distinctive spectral ranges: below and above $\approx1$ GHz. Below 1 GHz, the emission was much more intense than at higher frequencies (the maximum flux was at 610 MHz of about $10^4$~sfu); the polarization degree in this range was of about 38\,\%. Above 1 GHz, the maximum radio flux was observed around 9.4 GHz and was about 400 sfu; the polarization degree was low (less than 10\,\%). We conclude that the radio emission at low frequencies was likely produced by the plasma emission mechanism (due to strong radio flux and high polarization degree), while the emission above 1 GHz was generated by the GS mechanism. Therefore, below we analyze and simulate only the high-frequency part of the observed radio spectrum.

\begin{figure}
\centering
\includegraphics[width=1.0\linewidth]{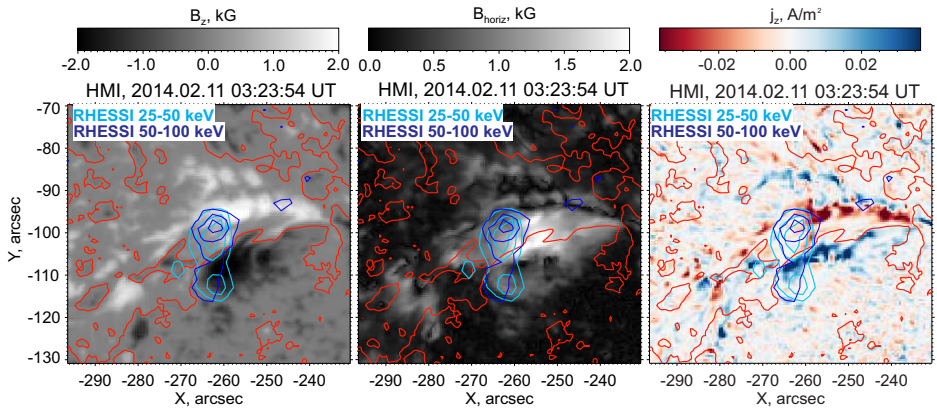}
\caption{Magnetogram of the selected active region. Left panel: the vertical magnetic field ($B_z$) map; middle panel: the horizontal magnetic field map; right panel: the vertical electric current ($j_z$) map. In all panels, the red contour marks the neutral line of the photospheric magnetic field ($z$-component). The RHESSI HXR contours in two energy ranges (for the time interval of 03:27:16\,--\,03:28:04 UT) are overplotted.}
\label{HMI}
\end{figure}

\subsection{Source structure}\label{SDO}
To study the magnetic field structure, we use observations of HMI, which provides vector magnetograms with 720 second cadence. We have selected the magnetogram closest to the flare impulsive phase (i.e. at 03:23:54~UT) and recalculated all of the magnetic field $\mathbfit{B}$. components from the local helioprojective Cartesian system to the Heliocentric Spherical coordinate system. The resulting vertical and horizontal components of the magnetic field at the photosphere level are shown in the left and middle panels of Figure \ref{HMI}, respectively. The vertical component of electric current is calculated as $j_z=[\nabla\times\mathbfit{B}]_z$ and is shown in the right panel of Figure \ref{HMI}. The neutral line (NL) of the photospheric magnetic field (the $B_z$ component) is marked by red color. The HMI vector magnetogram reprojected onto Heliocentric Spherical coordinate system is used also for magnetic field extrapolation as boundary conditions (see Section \ref{ModRad}).

In the same Figure, the RHESSI HXR contours in two energy ranges ($25-50$ and $50-100$ keV) are plotted. The HXR contours correspond to the time interval of 03:27:16\,--\,03:28:04 UT, i.e. around the flare peak; the HXR images were reconstructed using the CLEAN algorithm. The HXR maps reveal two strong sources corresponding to the footpoints of the flare loop located on both sides of the NL; the northern HXR source is more intense than the southern one. The double HXR emission sources are located in the vicinity of strong electric currents (up to 0.1 A $\textrm{m}^{-2}$), which indicates twisted magnetic flux tubes in the region where electrons are accelerated. The strongest horizontal photospheric magnetic field ($\approx 2$ kG) is detected near the NL between the HXR sources.

\begin{figure}
\centering
\includegraphics[width=1.0\linewidth]{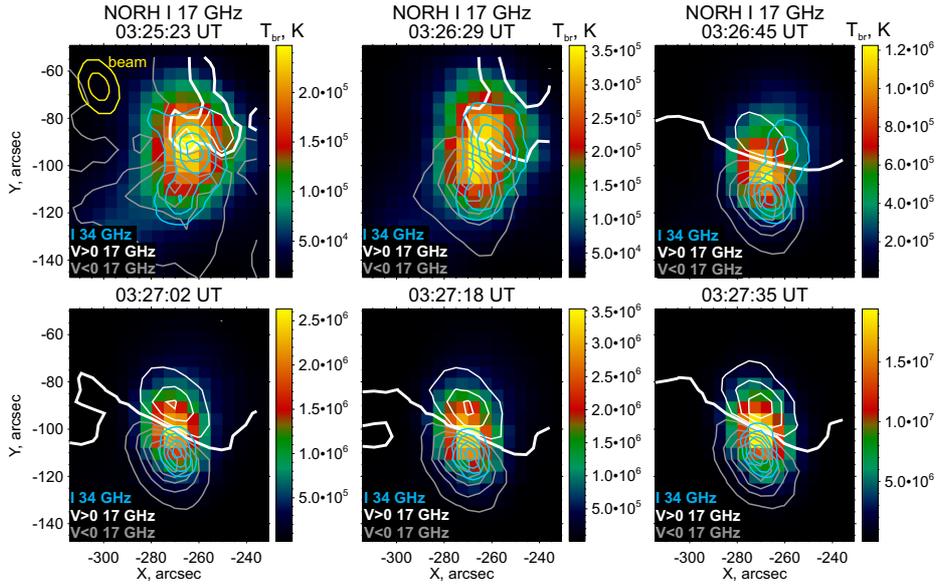}
\caption{NoRH microwave images of the 11 February, 2014 flare at different times. Colored background: 17 GHz Stokes $I$ map; blue contours: 34 GHz Stokes $I$ map. White and grey contours show negative and positive Stokes $V$ at 17 GHz. The polarization sign inversion line of the Stokes $V$ map at 17 GHz is shown by thick white line. The yellow contours show the NoRH beam shape (at half maximum level) at 17 (large ellipse) and 34 (small ellipse) GHz.}
\label{NORH}
\end{figure}

Figure \ref{NORH} shows the NoRH microwave images of the flare at six different times; the bottom right panel corresponds to the microwave emission maximum at 03:27:25 UT. We remind that NoRH produces microwave maps of the Sun at the frequencies of 17 GHz (Stokes $I$ and $V$) and 34 GHz (Stokes $I$ only) with spatial resolution of up to $10''$ and $5''$, respectively; the images were synthesized using the CLEAN algorithm.

It follows from the images that in the beginning of the impulsive phase, the polarization inversion was not very pronounced (first two panels in Figure \ref{NORH}) and the integral Stokes $V$ component was mostly negative. Radio maps at 34 GHz reveal a double emission source corresponding to the flare loop footpoints. Beginning from 03:27:00 UT, a clear PSIL separates two flare regions of the opposite polarizations (the negative Stokes $V$ component is still dominating) and only single 34 GHz emission source is detected, possibly corresponding to the flare loop top.

\begin{figure}
\centering
\includegraphics[width=1.0\linewidth]{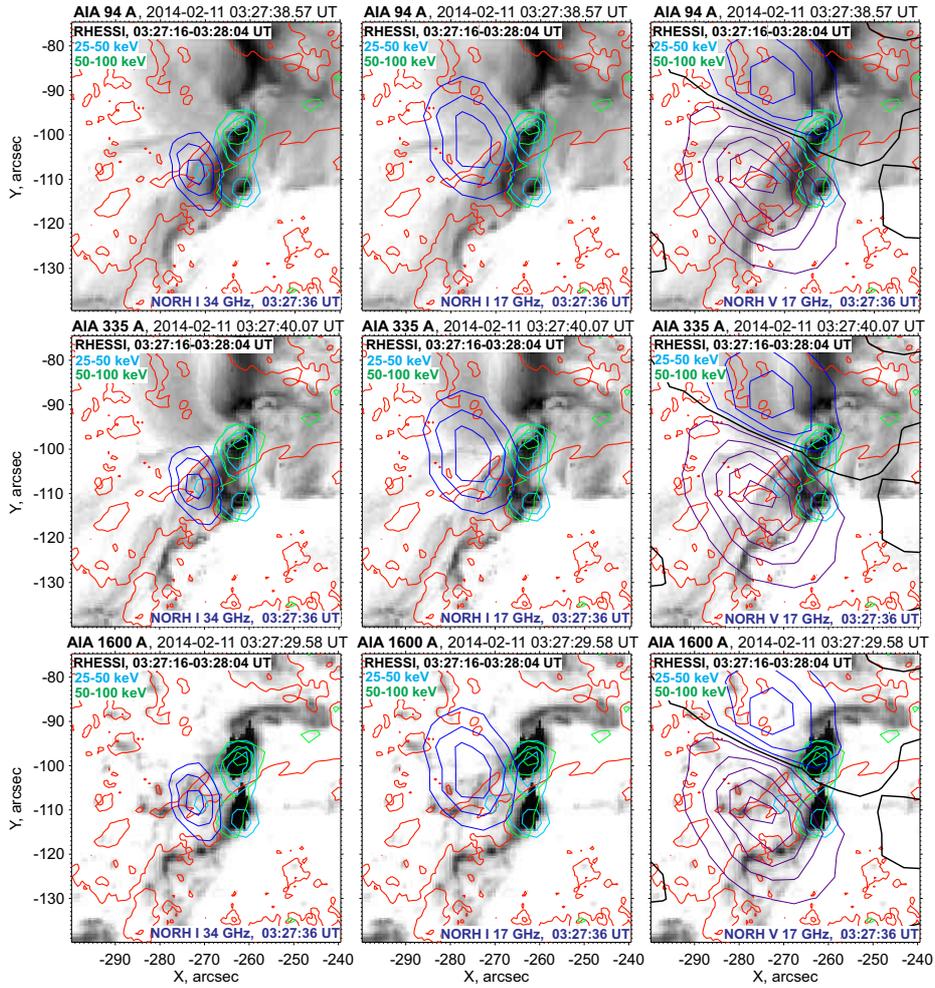}
\caption{Images of the 11 February, 2014 flare at different wavelengths. Background greyscale images: AIA 94 \AA{} (top row), 335 \AA{} (middle row), and 1600 \AA{} (bottom row); the neutral line is marked by red color. Cyan and green contours in the first three columns: RHESSI HXR maps at $25-50$ and $50-100$ keV, respectively (at 50, 70 and 90\% levels). NoRH microwave maps are shown by blue contours for 17 GHz Stokes $I$ in the first column, blue contours for 34 GHz Stokes $I$ in the second column, and blue and violet contours (for $V>0$ and $V<0$, respectively) for 17 GHz Stokes $V$ in the third column, with the polarization sign inversion line marked by the thick black line.}
\label{AIA}
\end{figure}

\begin{figure}
\centering
\includegraphics[width=1.0\linewidth]{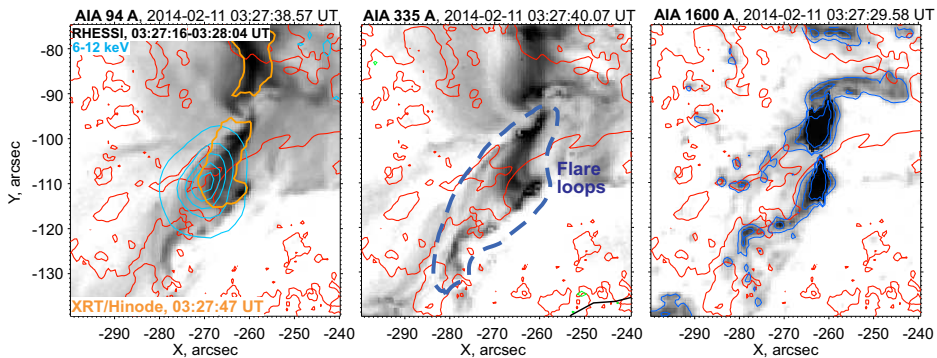}
\caption{Images of the 11 February, 2014 flare at different wavelengths. Background greyscale images: AIA 94 \AA{} (left), 335 \AA{} (middle), and 1600 \AA{} (right); the neutral line is marked by red color. Cyan contours in the left panel: RHESSI SXR map at $6-12$ keV (at 10, 30, 50, 70, and 90\,\% levels). Orange contour in the left panel: {\it Hinode} XRT X-ray map (at 90\,\% level). Dashed blue closed curve in the middle panel marks the flare loops system. Blue contours in the bottom right panel are drawn at the 10, 30, and 90\,\% levels of the AIA 1600 \AA{} emission.}
\label{AIA2}
\end{figure}

In Figure \ref{AIA}, we present the AIA images taken at 03:27:39 UT, i.e. near the flare peak. The EUV channels of 94 and 335 \AA{} (top and middle rows of Figure \ref{AIA}, respectively) were chosen because they experience the least saturation due to having the lowest sensitivities; these channels correspond to the emission of hot plasma with an average temperature of several MK. The UV channel of 1600 \AA{} (the bottom row) represents the continuum and C~{\sc iv} line emission produced in chromospheric regions. The AIA images in Figure \ref{AIA} are in logarithmic scale and thresholded to enhance contrast. In the same Figure, we overplot the RHESSI hard X-ray contours (in the energy ranges of $6-12$, $25-50$, and $50-100$ keV and in the same time interval as in Figure \ref{HMI}), the NoRH microwave contours (at the flare peak). RHESSI soft X-ray contours in the energy range of 6-12 keV and the {\it Hinode} XRT X-ray contours (near the flare peak) are shown in the left panel of Figure \ref{AIA2}.

In the AIA images one can see the loop system where the flare energy release occurred; this loop system is outlined schematically by the blue dashed contour in the middle panel of Figure \ref{AIA2}. The HXR emission sources coincide with the brightest pixels of the three AIA images considered. One can see that the loops connect flare ribbons that consist of small bright kernels with size as small as $2\times 10^{16}$ $\textrm{cm}^2$. The total area of the saturated UV 1600 \AA{} source (blue 90\,\% contour in the right panel of Figure \ref{AIA2}) is of about $3\times 10^{17}$ $\textrm{cm}^2$; below (in Section \ref{nonthermal}) we use this value as an estimation of the total cross-sectional area of the precipitating nonthermal electron beams. At the same time, the total area of the flare ribbons (10\,\% contour) is of about $2\times 10^{18}$ $\textrm{cm}^2$. The hottest loop in the arcade is detected by the XRT, which is sensitive to hot plasma at temperatures in the range of $1-10$ MK; this hottest loop is marked by the yellow contour in the left panel of Figure \ref{AIA2}. The single RHESSI SXR source ($6-12$ keV) also corresponds to the top of the X-ray loop-like bright structure observed by XRT.

The NoRH microwave sources correspond to the top of the flare loop system. One can also see that the weak HXR $25-50$ keV emission source coincides well with the 34 GHz microwave source. Strong emission at the loop top indicates accumulation of nonthermal electrons there (possibly connected with trapping).

Summarizing the observations, the imaging data presented in this section have allowed us to identify the flare loop system where the flare energy release occurred. We use these results for modelling of the flare magnetic structure with \textsf{GX Simulator} (see Section \ref{ModRad}). The RHESSI, NoRP, and RSTN data are used to determine the spectral parameters of accelerated electrons (see Section \ref{nonthermal}).

\section{X-ray and Radio Spectra}\label{nonthermal}
Figure \ref{spec} shows an example of RHESSI X-ray and composite NoRP and RSTN radio spectra of the flare considered; the spectra were recorded near the flare peak. Analysis of the spectra allows us to estimate various parameters of the thermal and nonthermal electrons. In particular, the X-ray spectrum (left panel of Figure \ref{spec}) was fitted with a superposition of a single-temperature bremsstrahlung radiation function and a double power-law to account for the thermal and nonthermal components, respectively. The break energy [$E_{\mathrm{low}}$] and the low-energy spectral index (at $E<E_{\mathrm{low}}$) of the nonthermal component were fixed to be 20 keV and 1.5, respectively, to improve the numerical stability and reduce the fitting errors; the value of $E_{\mathrm{low}}=20$ keV was chosen because it corresponds to the visible intersection of the thermal and nonthermal parts of the X-ray spectrum. As a result, we obtained the temperature [$T$] and emission measure [$EM$] of the thermal plasma, as well as the normalization factor [$A$] (i.e. the spectral flux density at $E=E_{\mathrm{low}}$) and the high-energy spectral index $\gamma$ (at $E>E_{\mathrm{low}}$) of the nonthermal X-ray component. The latter parameters allow us to estimate the total nonthermal X-ray photon flux above $E_{\mathrm{low}}$ as $I_{\mathrm{ph}}(E>E_{\mathrm{low}})=AE_{\mathrm{low}}/(\gamma-1)$.

The right panel of Figure \ref{spec} shows the radio spectrum around the flare peak combined from the NoRP and RSTN data. In particular, the grey line connecting two NoRP data points at 17 and 35 GHz approximates the slope of the high-frequency (optically thin) spectrum part; the corresponding spectral index is calculated as $\alpha = -\log(I_{17}/I_{35})/\log(17/35)$, where $I_{17}$ and $I_{35}$ are radio fluxes at the corresponding frequencies.

\begin{figure}
\centering
\includegraphics[width=9.9cm]{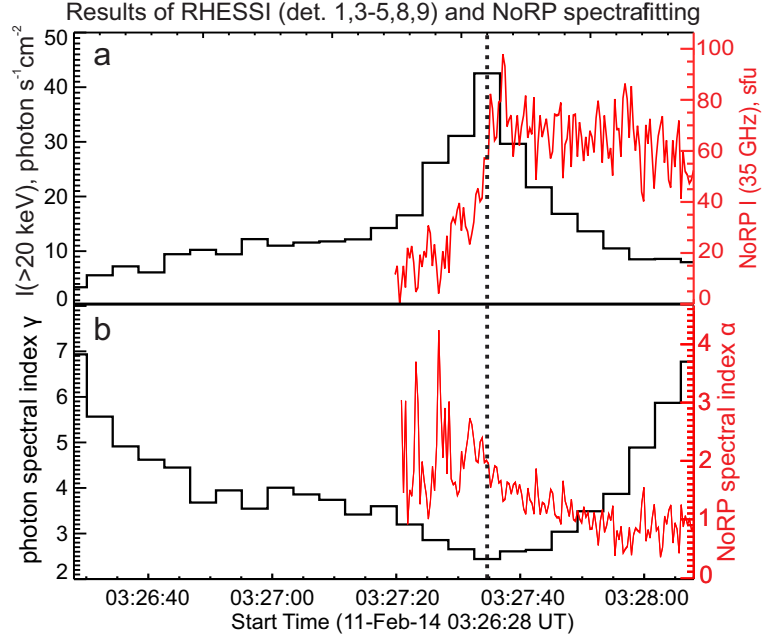}
\caption{Spectral parameters of the X-ray and microwave emissions during the flare impulsive phase. a) Integrated photon flux above~20 keV and microwave flux at 35 GHz (red). b) Spectral indices of HXR (black) and microwave (red) spectra. Vertical line marks the HXR maximum.}
\label{spec_fit}
\end{figure}

The temporal evolution of the emission spectral parameters is shown in Figure \ref{spec_fit}. The total HXR nonthermal photon flux above 20 keV and the radio flux at 35 GHz are shown in the top panel. In particular, the HXR spectrum varied according to the soft-hard-soft scenario (bottom panel); the hardest spectrum (with the spectral index of $\gamma\approx 2.44$) was observed at the HXR intensity peak. In contrast, the microwave spectral index near the flare peak and after it exhibited gradual hardening, which probably reflects the particle trapping process. Below, we consider the time interval around the flare peak to determine the model parameters for the \textsf{GX Simulator}.

In the thick-target approach, the X-ray photon spectrum is related to the nonthermal electron spectrum as \citealp{Syrovatskii1972} and \citealp{Brown1971}:
\begin{equation}
F(E>E_{\mathrm{low}})=1.02\times 10^{34}\frac{(\gamma-1)^2}{\beta(\gamma-1, 1/2)}\frac{I_{\mathrm{ph}}(E>E_{\mathrm{low}})}{E_{\mathrm{low}}},
\end{equation}
where $F({E>E_{\mathrm{low}}})$ [electrons $\textrm{s}^{-1}$] is the total nonthermal electon flux above $E_{\mathrm{low}}$, $I_{\mathrm{ph}}({E>E_{\mathrm{low}}})$ [photons $\textrm{s}^{-1}$ $\textrm{cm}^{-2}$] is the total photon flux above $E_{\mathrm{low}}$, $\beta(x, y)$ is the beta function, and $E_{\mathrm{low}}$ is in keV; the spectral index of accelerated electrons in the HXR source region is related to the emission spectral index as $\delta_{\mathrm{HXR}} =\gamma + 1$. In turn, the electron flux is related to the electron concentration $n_0$ (for a power-law spectrum, in the nonrelativistic approximation) as:
\begin{equation}
n_0(E>E_{\mathrm{low}}) = \frac{F(E>E_{\mathrm{low}})}{S} \sqrt{\frac{m_{\mathrm{e}}}{2E_{\mathrm{low}}}}\frac{\delta_{\mathrm{HXR}}-3/2}{\delta_{\mathrm{HXR}}-1},
\end{equation}
where $m_{\mathrm{e}}$ is the electron mass and $S$ is the precipitation area of the nonthermal electrons (all parameters in this formula are in CGS units). Using the HXR spectral fitting results near the flare peak, we estimate the total electron flux above 20 keV as $F(E>E_{\mathrm{low}})\simeq (2.6\pm 0.2)\times 10^{34}$ electrons $\textrm{s}^{-1}$. To estimate the area $S$, we use the AIA 1600 \AA{} images (see Section \ref{SDO}); at the impulsive phase of the flare, the total area of saturated bright UV emission sources was of about $S\simeq 3\times 10^{17}$ $\textrm{cm}^{-2}$. Therefore we estimate the concentration of nonthermal electrons (above 20 keV) near the flare peak as $n_0(E>E_{\mathrm{low}})\approx 10^7$ $\textrm{cm}^{-3}$.

We note that, according to \citealp{Dulk1985}, the spectral index of the optically thin gyrosynchrotron microwave emission [$\alpha$] is related to the spectral index of the microwave-emitting nonthermal electrons [$\delta_{\mathrm{MW}}$] as $\alpha=(\delta_{\mathrm{MW}}-1.22)/0.9$. For $\alpha=1.7$ (i.e. near the flare peak time), this implies $\delta_{\mathrm{MW}}=2.75$, which is harder than $\delta_{\mathrm{HXR}}=3.44$ deduced from the HXR spectral analysis. In the simulations below, we adopt the electron spectral index determined from the microwave observations (i.e. $\delta=\delta_{\mathrm{MW}}$), because it reflects directly the characteristics of nonthermal electrons inside the coronal flaring loops. On the other hand, for the concentration of nonthermal electrons we use the estimations based on the HXR observations, as the best approach available. The estimated parameters of the spectrum of nonthermal electrons are imported to \textsf{GX Simulator} to model the radio emission Stokes $I$ and $V$ maps at the frequencies of 17 and 34 GHz.

\section{Modelling of Radio Emission}
\label{ModRad}

\subsection{Selection of the Magnetic Loops}

3D modelling of the microwave radio emission is made using the \texttt{GX Simulator} package \citep{Nita2015}. This IDL-based program is an interactive tool allowing one to select ``active'' magnetic structures in the solar corona; then, using analytical expressions, one defines thermal and nonthermal particle distributions along and across the selected magnetic structures. The Stokes $I$ and $V$ microwave maps are calculated using the \textsf{Fast Gyrosynchrotron Codes} \citep{Fleishman2010} where the authors used some analytical approaches and numerical methods to calculate the microwave emission with high speed and good accuracy for different energy and pitch-angle distributions of nonthermal particles. In our modelling, the main task is to explain the emission at 17 and 34 GHz as we have imaging data for these frequencies. However, we will discuss the emission at lower frequencies as well.

\begin{figure}
\centering
\includegraphics[width=1.0\linewidth]{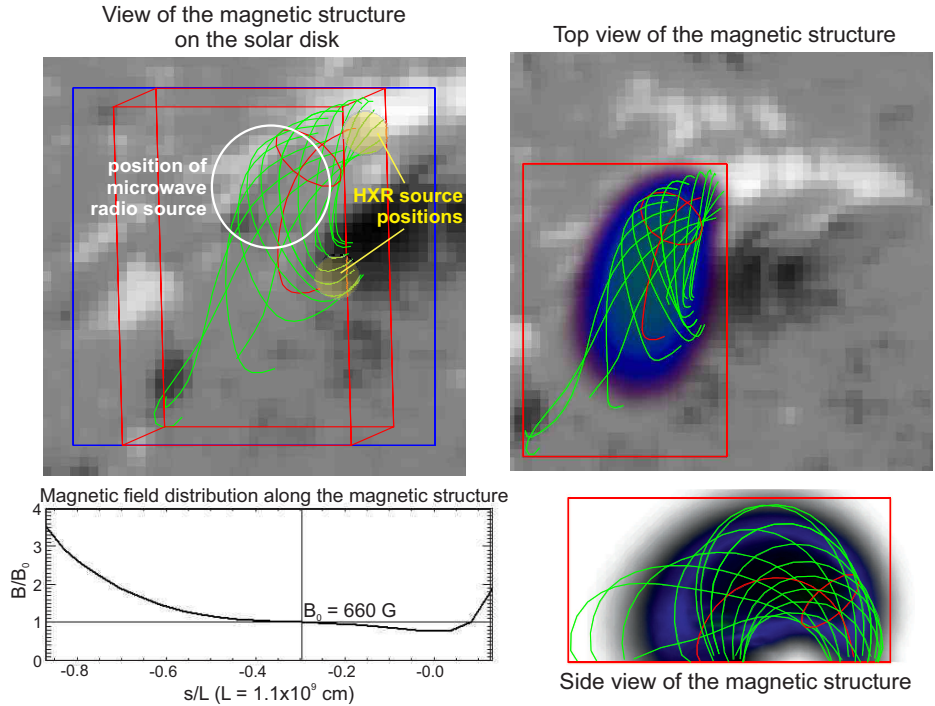}
\caption{The magnetic loop selected to model the microwave emission; the magnetic field was reconstructed using the NLFFF approach. The actual orientation of the magnetic loop (as seen from the Earth) is shown in the top left panel. Top and side views of the magnetic loop with the model distribution of nonthermal electrons inside it (shown by blue color) are presented in top right and bottom right panels, respectively. Bottom left panel: distribution of the magnetic field along the loop axis.}
\label{scheme1}
\end{figure}

As described above (Section \ref{SDO}), from the multiwavelength observations we have identified the magnetic structure where the flare energy release took place. To reconstruct the 3D distribution of the coronal magnetic field, we use nonlinear force-free (NLFFF) magnetic field extrapolation \citep{Wheatland2000} with the SDO HMI vector magnetogram used as boundary condition. The extrapolation is made using the optimization algorithm (implemented by \citealp{Rudenko2009}), which transforms the initial potential state of the magnetic field to the force-free one, according to the photosphere vector magnetogram provided. The extrapolation results are shown in Figure \ref{scheme1}, where the field lines of the selected magnetic loop are plotted. The magnetic field lines were chosen to reproduce the observed locations of the HXR footpoints and coronal microwave source. The noticeably twisted magnetic structure is elongated along the NL and reproduces nicely the shape of the observed EUV flare structure. The magnetic field strengths in the northern and southern footpoints are of about 2300 and 1200, respectively. The minimum value of the magnetic field strength within the loop is of about 600 G.

\begin{figure}
\centering
\includegraphics[width=1.0\linewidth]{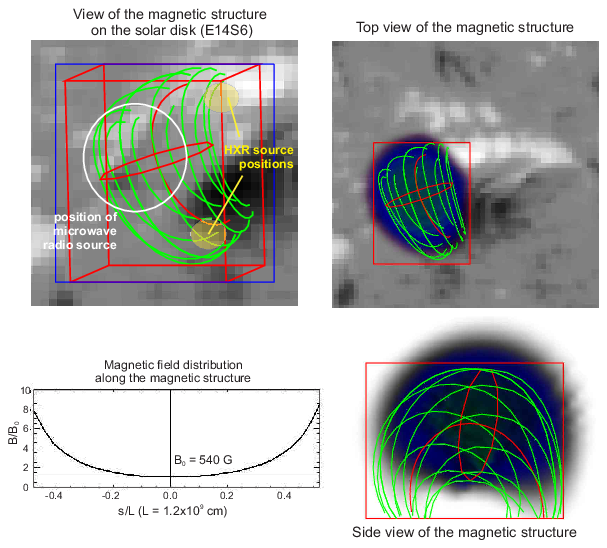}
\caption{The magnetic loop selected to model the microwave emission; the magnetic field was reconstructed using the potential field approach. The actual orientation of the magnetic loop (as seen from the Earth) is shown in the top left panel. Top and side views of the magnetic loop with the model distribution of nonthermal electrons inside it (shown by blue color) are presented in top right and bottom right panels, respectively. Bottom left panel: distribution of the magnetic field along the loop axis.}
\label{scheme3}
\end{figure}

To analyze the effect of the magnetic structure, we have also tried to interpret the observations using the potential magnetic field approach, which assumes no electric currents in the considered volume. The resulting magnetic field lines are shown in Figure \ref{scheme3}; the magnetic field strength in the footpoints is about 1300 G, while the minimum value of the field strength (at the loop top) is about 160 G. One can see that potential magnetic field cannot explain the observed shape of the flare magnetic structure. Nevertheless, below we compare the results of radio emission modelling for these two magnetic field extrapolations with the NoRH observations and demonstrate that the NLFFF approach reproduces the polarization pattern much better.

\subsection{Source Parameters}
Nonthermal electrons are non-uniformly distributed inside the magnetic structure. We consider a Gaussian shape of the distributions of the nonthermal electrons along and across the magnetic structure. The distributions are determined by the expressions $n(r,l) = n_0\exp{[-(0.75r/R)^2]}\exp{[-(2.5l/L)^2]}$ for the NLFFF model and $n(r,l) = n_0\exp{[-(0.75r/R)^2]}\exp{[-(4l/L)^2]}$ for the potential magnetic field model, respectively; here $R$ is the loop radius, $r$ is the coordinate across the loop, $L$ is the loop length, and $l$ is the coordinate along the loop. The peak number density is chosen to be $n_0=10^7$ $\textrm{cm}^{-3}$ at the loop top. These shapes were chosen to reproduce the loop-top position of the observed radio source and obtain a sufficient radio flux to fit the observed radio spectrum. The energy spectrum of nonthermal electrons is described by power-law function with spectral index $\delta=2.75$, the low-energy cutoff $E_{\mathrm{low}}=20$ keV, and the high-energy cutoff $E_{\mathrm{high}}=10$ MeV; we consider an isotropic pitch-angle distribution of the nonthermal electrons. The background thermal plasma is uniformly distributed in the loop and has the number density of $n_{\mathrm{th}}=5\times 10^{10}$ $\textrm{cm}^{-3}$ and temperature of $T=23.2$ MK.

We consider two regimes of microwave polarization transfer, referred to as the ``strong'' and ``exact'' mode coupling approaches. In the strong coupling approach, the emission polarization sign is determined by the magnetic field direction in the emission source (i.e. neglecting any propagation effects), while in the exact coupling approach, possible polarization changes during propagation (in the quasi-transverse magnetic field regions) are taken into account \citep{Cohen1960,Zheleznyakov1964}; these approaches are implemented in the \textsf{GX Simulator}. The emission intensity remains nearly the same in both cases.

\begin{figure}
\centering
\includegraphics[width=1.0\linewidth]{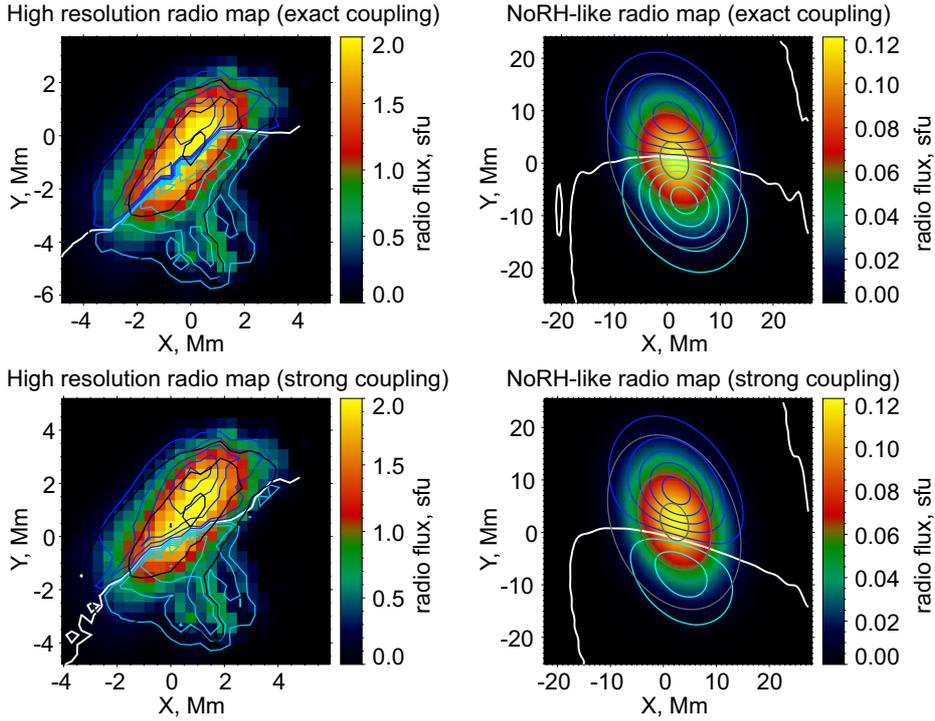}
\caption{Simulated microwave images for the model based on the NLFFF magnetic field extrapolation (see Figure \protect\ref{scheme1}). The Stokes $I$ maps at 17 and 34 GHz are shown by color background and white contours, respectively; the Stokes $V$ maps at 17 GHz are shown by cyan and blue contours for $V<0$ and $V>0$, respectively. The PSIL is marked by thick white line. Left panels: original (high-resolution) simulated radio maps; right panels: the simulated radio maps convolved with the NoRH beam. Top panels: exact mode coupling; bottom panels: strong mode coupling.}
\label{res1}
\end{figure}

\begin{figure}
\centering
\includegraphics[width=1.0\linewidth]{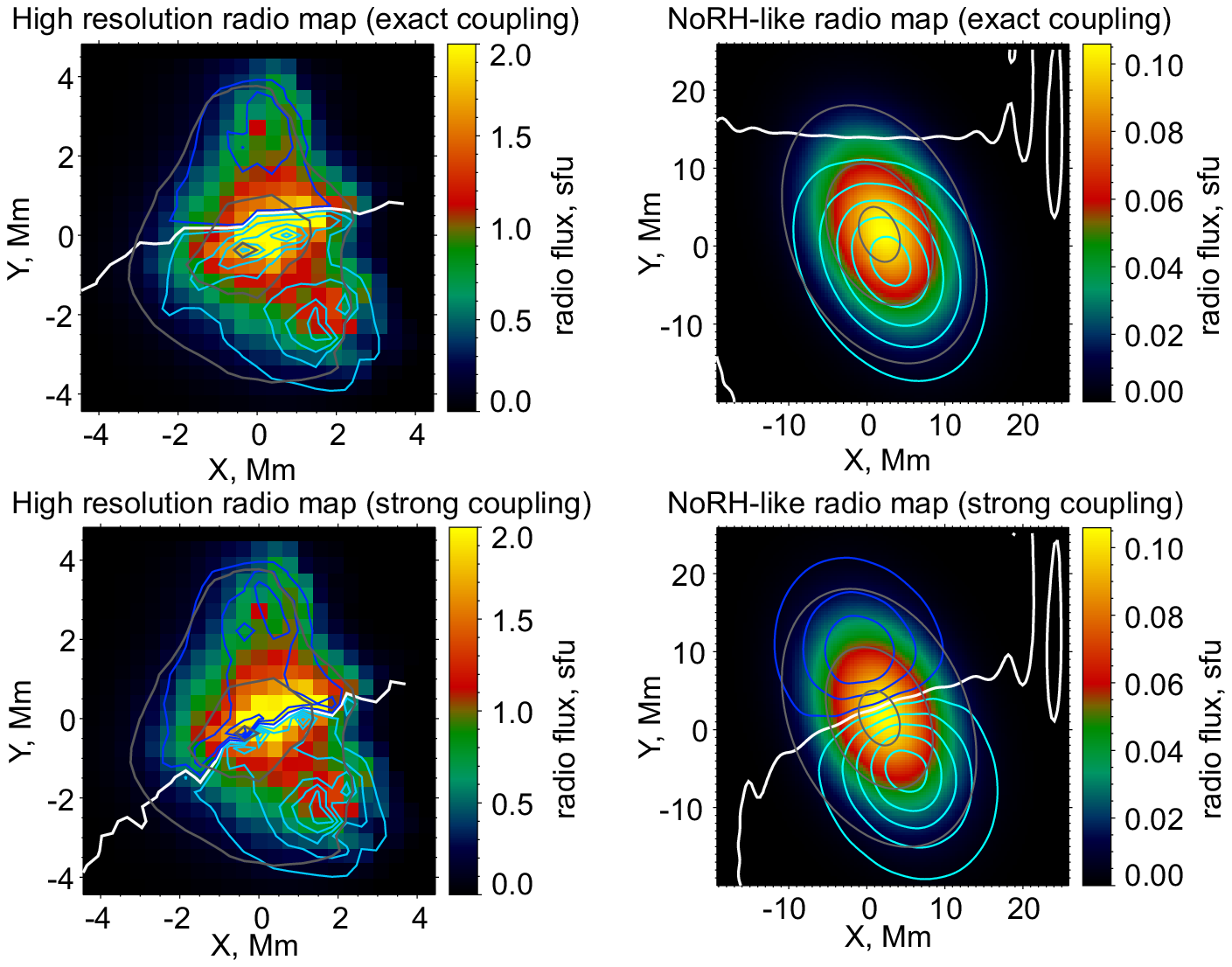}
\caption{Simulated microwave images for the model based on the potential field extrapolation (see Figure \protect\ref{scheme3}). The Stokes $I$ maps at 17 and 34 GHz are shown by color background and white contours, respectively; the Stokes $V$ maps at 17 GHz are shown by cyan and blue contours for $V<0$ and $V>0$, respectively. The PSIL is marked by thick white line. Left panels: original (high-resolution) simulated radio maps; right panels: the simulated radio maps convolved with the NoRH beam. Top panels: exact mode coupling; bottom panels: strong mode coupling.}
\label{res3}
\end{figure}

\begin{figure}
\centering
\includegraphics[width=1.0\linewidth]{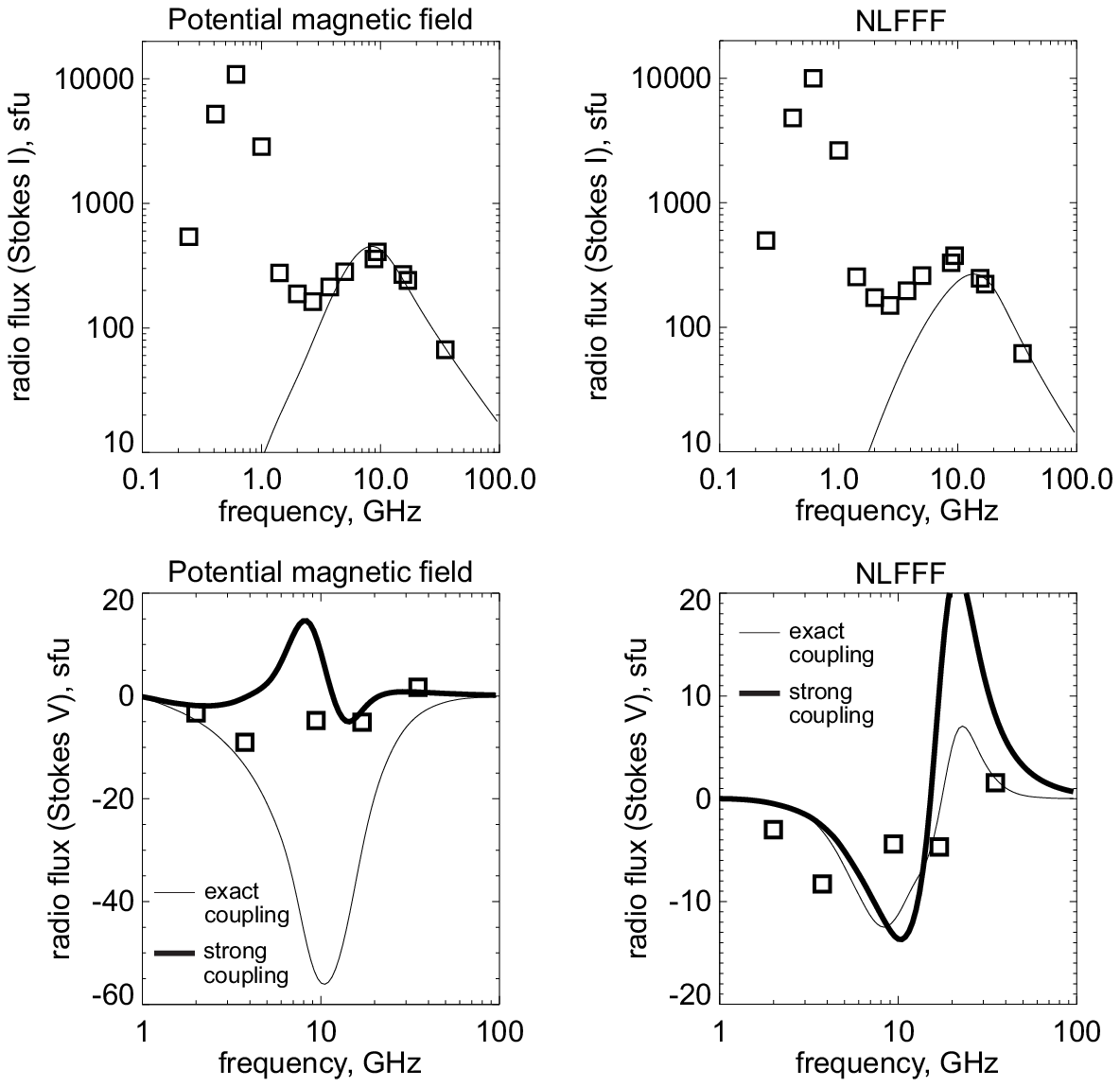}
\caption{The observed (squares) and simulated (solid lines) microwave spectra. Left panels: the model based on the potential magnetic field extrapolation; right panel: the model based on the NLFFF extrapolation. Top panels: emission intensity (Stokes $I$, NoRP and RSTN observations are shown); bottom panels: circular polarization (Stokes $V$, NoRP data only). In the bottom panels, thick and thin lines correspond to strong and exact mode coupling regimes, respectively.}
\label{spec_mod}
\end{figure}

\subsection{Simulation Results}
The simulated radio images are presented in Figures \ref{res1}--\ref{res3} for the NLFFF and potential magnetic field extrapolations, respectively; we demonstrate both the original (high-resolution) simulated radio maps and those maps convolved with the NoRH response function. The total (spatially integrated) emission spectra are shown in Figure \ref{spec_mod}.

For the radio emission modelling based on the NLFFF magnetic field extrapolation, one can see that the observed intensity spectrum (top right panel of Figure \ref{spec_mod}) is nicely fitted by the simulated spectrum at the frequencies of 17 and 34 GHz. The observed radio emission at low frequencies ($<10$ GHz) is characterized by higher fluxes compared with the simulated ones; the simulated peak frequency is slightly higher than that determined from observations. The model Stokes $V$ spectrum (bottom right panel of Figure \ref{spec_mod}) is close to the observational data points in the case of exact mode coupling, while the strong mode coupling model provides slightly worse agreement with the observations. Also, we can see that the observed radio maps presented in the previous sections are nicely explained by the simulated ones (Figure \ref{res1}). The most important factor is that the simulated PSIL has an orientation very similar to the observed one. In the case of the exact mode coupling, the ratio of the maxima of negative and positive Stokes $V$ sources is very close to the observed one, while the strong mode coupling regime does not fit the observations.

Figure \ref{res3} and left panel of Figure \ref{spec_mod} present results of radio emission modelling based on the potential magnetic field extrapolation. The simulated intensity spectrum (top left panel of Figure \ref{spec_mod}) fits the observed data points very well; on the other hand, the polarization spectrum (bottom left panel of Figure \ref{spec_mod}) is reproduced less accurately with both the strong and exact mode coupling models. The simulated radio maps (Figure \ref{res3}) reveal significant differences between the observed and simulated polarization distributions: for the strong mode coupling regime, the simulated PSIL has the inclination opposite to that in the observed NoRH Stokes $V$ map, while in the case of exact mode coupling the simulated polarization is mostly negative, which also contradicts the observations.

Thus, we have shown that the potential magnetic field configuration cannot explain the observed polarization of radio emission at 17 GHz. However, the total intensity spectrum is explained by this approach very well. This may be connected with the fact that potential magnetic field has a tendency to decrease rapidly with height. As a result, nonthermal electrons within potential magnetic structures occupy a large volume with weak magnetic field and hence the gyrosynchrotron emission at lower frequencies is more pronounced compared with the emission from the footpoints with stronger magnetic field; the peak frequency is shifted towards lower frequencies. Another factor contributing to the higher intensity at lower frequencies in the potential magnetic configurations is a larger area of the loop-top region. In contrast, the NLFFF magnetic field structure is characterised by smaller volume and stronger magnetic field; that is why the simulated spectrum fails to explain the low-frequency spectral part of the observed one. We therefore suggest that a perfect agreement between the simulations and the observations might be achieved by introducing a hybrid model consisting of a compact twisted magnetic structure surrounded by a  quasipotential shell with lower magnetic field; however, we do not analyze such a model in details as there were no imaging data at lower frequencies to compare with simulations.

\section{Conclusions}
We analyzed the solar flare on 11~February, 2014. We found that the radio emission source with polarization inversion was located inside the twisted magnetic structure. Three dimensional modelling of the flare radio emission using \textsf{GX Simulator} showed that the observed polarization map can be explained only by assuming nonthermal electrons propagating in a twisted magnetic structure that can be reconstructed using the NLFFF magnetic field extrapolation technique; radio simulations based on the potential magnetic field cannot explain the observed polarization pattern. Thus, we demonstrated that polarization radio maps can be used for diagnosing the topology of the coronal magnetic structures filled with nonthermal electrons. However, interpretation of polarization maps is still limited by insufficient spatial resolution of the existing instruments (such as the Nobeyama Radioheliograph). New polarimetric observations with higher spatial resolution are needed; very perspective instruments for the proposed studies are the Siberian Radioheliograph (SRH, resolution $5''$ at 24 GHz), expanded Owens Valley Solar Array (eOVSA, resolution $3''$ at 17 GHz) and Mingantu Ultrawide Spectral Radioheliograph (MUSER, resolution $1.4''$ at 15 GHz).

\begin{acks}
This work was supported by the Russian Foundation of Basic Research (RFBR, grant 16-32-50172). A.A.~Kuznetsov acknowledges partial support from the RFBR grants 15-02-03717 and 15-02-03835. I.I.~Myshyakov acknowledges partial support from the RFBR grant 16-32-00315. I.N.~Sharykin is grateful to the colleagues from the Institute of Solar-Terrestrial Physics for discussions and hospitality during working at the ISTP.
\end{acks}

\section*{Disclosure of Potential Conflicts of Interest}
The authors claim that they have no conflicts of interest.


\bibliographystyle{spr-mp-sola}

\begin{thebibliography}{32}
\ifx\bisbn     \undefined \def\bisbn  #1{ISBN #1}\fi
\ifx\binits    \undefined \def\binits#1{#1}\fi
\ifx\bauthor   \undefined \def\bauthor#1{#1}\fi
\ifx\batitle   \undefined \def\batitle#1{#1}\fi
\ifx\bjtitle   \undefined \def\bjtitle#1{\textit{#1}}\fi
\ifx\bvolume   \undefined \def\bvolume#1{\textbf{#1}}\fi
\ifx\byear     \undefined \def\byear#1{#1}\fi
\ifx\bissue    \undefined \def\bissue#1{#1}\fi
\ifx\bfpage    \undefined \def\bfpage#1{#1}\fi
\ifx\blpage    \undefined \def\blpage #1{#1}\fi
\ifx\burl      \undefined \def\burl#1{\textsf{#1}}\fi
\ifx\href      \undefined \def\href#1#2{\textsf{#2}}\fi
\ifx\betal     \undefined \def\betal{\textit{et al.}}\fi
\ifx\bctitle   \undefined \def\bctitle#1{#1}\fi
\ifx\beditor   \undefined \def\beditor#1{#1}\fi
\ifx\bbtitle   \undefined \def\bbtitle#1{\textit{#1}}\fi
\ifx\bedition  \undefined \def\bedition#1{#1}\fi
\ifx\bseriesno \undefined \def\bseriesno#1{\textbf{#1}}\fi
\ifx\blocation \undefined \def\blocation#1{#1}\fi
\ifx\bsertitle \undefined \def\bsertitle#1{\textit{#1}}\fi
\ifx\bsnm      \undefined \def\bsnm#1{#1}\fi
\ifx\bsuffix   \undefined \def\bsuffix#1{#1}\fi
\ifx\bparticle \undefined \def\bparticle#1{#1}\fi
\ifx\barticle  \undefined \def\barticle#1{}\fi
\ifx\binstitute  \undefined \def\binstitute#1{#1}\fi
\ifx\bpublisher  \undefined \def\bpublisher#1{#1}\fi
\ifx\doiurl    \undefined
  \def\doiurl#1{\href{http://dx.doi.org/#1}{\textsf{DOI}}}\fi
\ifx\arxivurl  \undefined
  \def\arxivurl#1{\href{http://arxiv.org/abs/#1}{\textsf{arXiv}}}\fi
\ifx\adsurl    \undefined
  \def\adsurl#1{\href{http://adsabs.harvard.edu/abs/#1}{\textsf{ADS}}}\fi
\ifx\botherref \undefined \def\botherref#1{}\fi
\ifx\url       \undefined \def\url#1{\textsf{#1}}\fi
\ifx\bchapter  \undefined \def\bchapter#1{}\fi
\ifx\bbook     \undefined \def\bbook#1{}\fi
\ifx\bcomment  \undefined \def\bcomment#1{#1}\fi
\ifx\oauthor   \undefined \def\oauthor#1{#1}\fi
\ifx\citeauthoryear \undefined\def \citeauthoryear#1{#1}\fi
\ifx\endbibitem\undefined \def\endbibitem{}\fi
\ifx\bconflocation  \undefined \def\bconflocation#1{#1} \fi

\bibitem[\protect\citeauthoryear{{Battaglia}, {Grigis}, and
  {Benz}}{2005}]{Battaglia2005}
\begin{barticle}
\bauthor{\bsnm{{Battaglia}}, \binits{M.}},
\bauthor{\bsnm{{Grigis}}, \binits{P.C.}},
\bauthor{\bsnm{{Benz}}, \binits{A.O.}}:
\byear{2005},
\batitle{{Size dependence of solar X-ray flare properties}}.
\bjtitle{\aap}
\bvolume{439},
\bfpage{737}.
\doiurl{10.1051/0004-6361:20053027}.
\adsurl{2005A\%26A...439..737B}.
\end{barticle}
\endbibitem

\bibitem[\protect\citeauthoryear{{Brown}}{1971}]{Brown1971}
\begin{barticle}
\bauthor{\bsnm{{Brown}}, \binits{J.C.}}:
\byear{1971},
\batitle{{The Deduction of Energy Spectra of Non-Thermal Electrons in Flares
  from the Observed Dynamic Spectra of Hard X-Ray Bursts}}.
\bjtitle{\solphys}
\bvolume{18},
\bfpage{489}.
\doiurl{10.1007/BF00149070}.
\adsurl{1971SoPh...18..489B}.
\end{barticle}
\endbibitem

\bibitem[\protect\citeauthoryear{{Cohen}}{1960}]{Cohen1960}
\begin{barticle}
\bauthor{\bsnm{{Cohen}}, \binits{M.H.}}:
\byear{1960},
\batitle{{Magnetoionic Mode Coupling at High Frequencies.}}
\bjtitle{\apj}
\bvolume{131},
\bfpage{664}.
\doiurl{10.1086/146878}.
\adsurl{1960ApJ...131..664C}.
\end{barticle}
\endbibitem

\bibitem[\protect\citeauthoryear{{D{\'e}moulin}, {Priest}, and
  {Lonie}}{1996}]{Demoulin1996}
\begin{barticle}
\bauthor{\bsnm{{D{\'e}moulin}}, \binits{P.}},
\bauthor{\bsnm{{Priest}}, \binits{E.R.}},
\bauthor{\bsnm{{Lonie}}, \binits{D.P.}}:
\byear{1996},
\batitle{{Three-dimensional magnetic reconnection without null points 2.
  Application to twisted flux tubes}}.
\bjtitle{\jgr}
\bvolume{101},
\bfpage{7631}.
\doiurl{10.1029/95JA03558}.
\adsurl{1996JGR...101.7631D}.
\end{barticle}
\endbibitem

\bibitem[\protect\citeauthoryear{{Dulk}}{1985}]{Dulk1985}
\begin{barticle}
\bauthor{\bsnm{{Dulk}}, \binits{G.A.}}:
\byear{1985},
\batitle{{Radio emission from the sun and stars}}.
\bjtitle{\araa}
\bvolume{23},
\bfpage{169}.
\doiurl{10.1146/annurev.aa.23.090185.001125}.
\adsurl{1985ARA\%26A..23..169D}.
\end{barticle}
\endbibitem

\bibitem[\protect\citeauthoryear{{Fleishman} and
  {Kuznetsov}}{2010}]{Fleishman2010}
\begin{barticle}
\bauthor{\bsnm{{Fleishman}}, \binits{G.D.}},
\bauthor{\bsnm{{Kuznetsov}}, \binits{A.A.}}:
\byear{2010},
\batitle{{Fast Gyrosynchrotron Codes}}.
\bjtitle{\apj}
\bvolume{721},
\bfpage{1127}.
\doiurl{10.1088/0004-637X/721/2/1127}.
\adsurl{2010ApJ...721.1127F}.
\end{barticle}
\endbibitem

\bibitem[\protect\citeauthoryear{{Golub} \textit{et~al.}}{2007}]{Golub07}
\begin{barticle}
\bauthor{\bsnm{{Golub}}, \binits{L.}},
\bauthor{\bsnm{{Deluca}}, \binits{E.}},
\bauthor{\bsnm{{Austin}}, \binits{G.}},
\bauthor{\bsnm{{Bookbinder}}, \binits{J.}},
\bauthor{\bsnm{{Caldwell}}, \binits{D.}},
\bauthor{\bsnm{{Cheimets}}, \binits{P.}},
\bauthor{\bsnm{{Cirtain}}, \binits{J.}},
\bauthor{\bsnm{{Cosmo}}, \binits{M.}},
\bauthor{\bsnm{{Reid}}, \binits{P.}},
\bauthor{\bsnm{{Sette}}, \binits{A.}},
\bauthor{\bsnm{{Weber}}, \binits{M.}},
\bauthor{\bsnm{{Sakao}}, \binits{T.}},
\bauthor{\bsnm{{Kano}}, \binits{R.}},
\bauthor{\bsnm{{Shibasaki}}, \binits{K.}},
\bauthor{\bsnm{{Hara}}, \binits{H.}},
\bauthor{\bsnm{{Tsuneta}}, \binits{S.}},
\bauthor{\bsnm{{Kumagai}}, \binits{K.}},
\bauthor{\bsnm{{Tamura}}, \binits{T.}},
\bauthor{\bsnm{{Shimojo}}, \binits{M.}},
\bauthor{\bsnm{{McCracken}}, \binits{J.}},
\bauthor{\bsnm{{Carpenter}}, \binits{J.}},
\bauthor{\bsnm{{Haight}}, \binits{H.}},
\bauthor{\bsnm{{Siler}}, \binits{R.}},
\bauthor{\bsnm{{Wright}}, \binits{E.}},
\bauthor{\bsnm{{Tucker}}, \binits{J.}},
\bauthor{\bsnm{{Rutledge}}, \binits{H.}},
\bauthor{\bsnm{{Barbera}}, \binits{M.}},
\bauthor{\bsnm{{Peres}}, \binits{G.}},
\bauthor{\bsnm{{Varisco}}, \binits{S.}}:
\byear{2007},
\batitle{{The X-Ray Telescope (XRT) for the Hinode Mission}}.
\bjtitle{\solphys}
\bvolume{243},
\bfpage{63}.
\doiurl{10.1007/s11207-007-0182-1}.
\adsurl{2007SoPh..243...63G}.
\end{barticle}
\endbibitem

\bibitem[\protect\citeauthoryear{{Gordovskyy} and
  {Browning}}{2011}]{Gordovskyy2011}
\begin{barticle}
\bauthor{\bsnm{{Gordovskyy}}, \binits{M.}},
\bauthor{\bsnm{{Browning}}, \binits{P.K.}}:
\byear{2011},
\batitle{{Particle Acceleration by Magnetic Reconnection in a Twisted Coronal
  Loop}}.
\bjtitle{\apj}
\bvolume{729},
\bfpage{101}.
\doiurl{10.1088/0004-637X/729/2/101}.
\adsurl{2011ApJ...729..101G}.
\end{barticle}
\endbibitem

\bibitem[\protect\citeauthoryear{{Gordovskyy} and
  {Browning}}{2012}]{Gordovskyy2012}
\begin{barticle}
\bauthor{\bsnm{{Gordovskyy}}, \binits{M.}},
\bauthor{\bsnm{{Browning}}, \binits{P.K.}}:
\byear{2012},
\batitle{{Magnetic Relaxation and Particle Acceleration in a Flaring Twisted
  Coronal Loop}}.
\bjtitle{\solphys}
\bvolume{277},
\bfpage{299}.
\doiurl{10.1007/s11207-011-9900-9}.
\adsurl{2012SoPh..277..299G}.
\end{barticle}
\endbibitem

\bibitem[\protect\citeauthoryear{{Gordovskyy}, {Browning}, and
  {Kontar}}{2017}]{Gordovskyy2017}
\begin{barticle}
\bauthor{\bsnm{{Gordovskyy}}, \binits{M.}},
\bauthor{\bsnm{{Browning}}, \binits{P.K.}},
\bauthor{\bsnm{{Kontar}}, \binits{E.P.}}:
\byear{2017},
\batitle{{Polarisation of microwave emission from reconnecting twisted coronal
  loops}}.
\bjtitle{\aap}
\bvolume{604},
\bfpage{A116}.
\doiurl{10.1051/0004-6361/201629334}.
\adsurl{2017A\%26A...604A.116G}.
\end{barticle}
\endbibitem

\bibitem[\protect\citeauthoryear{{Gordovskyy}
  \textit{et~al.}}{2013}]{Gordovskyy2013}
\begin{barticle}
\bauthor{\bsnm{{Gordovskyy}}, \binits{M.}},
\bauthor{\bsnm{{Browning}}, \binits{P.K.}},
\bauthor{\bsnm{{Kontar}}, \binits{E.P.}},
\bauthor{\bsnm{{Bian}}, \binits{N.H.}}:
\byear{2013},
\batitle{{Effect of Collisions and Magnetic Convergence on Electron
  Acceleration and Transport in Reconnecting Twisted Solar Flare Loops}}.
\bjtitle{\solphys}
\bvolume{284},
\bfpage{489}.
\doiurl{10.1007/s11207-012-0124-4}.
\adsurl{2013SoPh..284..489G}.
\end{barticle}
\endbibitem

\bibitem[\protect\citeauthoryear{{Gordovskyy}
  \textit{et~al.}}{2014}]{Gordovskyy2014}
\begin{barticle}
\bauthor{\bsnm{{Gordovskyy}}, \binits{M.}},
\bauthor{\bsnm{{Browning}}, \binits{P.K.}},
\bauthor{\bsnm{{Kontar}}, \binits{E.P.}},
\bauthor{\bsnm{{Bian}}, \binits{N.H.}}:
\byear{2014},
\batitle{{Particle acceleration and transport in reconnecting twisted loops in
  a stratified atmosphere}}.
\bjtitle{\aap}
\bvolume{561},
\bfpage{A72}.
\doiurl{10.1051/0004-6361/201321715}.
\adsurl{2014A\%26A...561A..72G}.
\end{barticle}
\endbibitem

\bibitem[\protect\citeauthoryear{{Guo} \textit{et~al.}}{2012}]{Guo2012}
\begin{barticle}
\bauthor{\bsnm{{Guo}}, \binits{J.}},
\bauthor{\bsnm{{Emslie}}, \binits{A.G.}},
\bauthor{\bsnm{{Massone}}, \binits{A.M.}},
\bauthor{\bsnm{{Piana}}, \binits{M.}}:
\byear{2012},
\batitle{{Properties of the Acceleration Regions in Several Loop-structured
  Solar Flares}}.
\bjtitle{\apj}
\bvolume{755},
\bfpage{32}.
\doiurl{10.1088/0004-637X/755/1/32}.
\adsurl{2012ApJ...755...32G}.
\end{barticle}
\endbibitem

\bibitem[\protect\citeauthoryear{{Hirayama}}{1974}]{Hirayama1974}
\begin{barticle}
\bauthor{\bsnm{{Hirayama}}, \binits{T.}}:
\byear{1974},
\batitle{{Theoretical Model of Flares and Prominences. I: Evaporating Flare
  Model}}.
\bjtitle{\solphys}
\bvolume{34},
\bfpage{323}.
\doiurl{10.1007/BF00153671}.
\adsurl{1974SoPh...34..323H}.
\end{barticle}
\endbibitem

\bibitem[\protect\citeauthoryear{{Jiang} \textit{et~al.}}{2006}]{Jiang2006}
\begin{barticle}
\bauthor{\bsnm{{Jiang}}, \binits{Y.W.}},
\bauthor{\bsnm{{Liu}}, \binits{S.}},
\bauthor{\bsnm{{Liu}}, \binits{W.}},
\bauthor{\bsnm{{Petrosian}}, \binits{V.}}:
\byear{2006},
\batitle{{Evolution of the Loop-Top Source of Solar Flares: Heating and Cooling
  Processes}}.
\bjtitle{\apj}
\bvolume{638},
\bfpage{1140}.
\doiurl{10.1086/498863}.
\adsurl{2006ApJ...638.1140J}.
\end{barticle}
\endbibitem

\bibitem[\protect\citeauthoryear{{Kupriyanova}
  \textit{et~al.}}{2010}]{Kupriyanova2010}
\begin{barticle}
\bauthor{\bsnm{{Kupriyanova}}, \binits{E.G.}},
\bauthor{\bsnm{{Melnikov}}, \binits{V.F.}},
\bauthor{\bsnm{{Nakariakov}}, \binits{V.M.}},
\bauthor{\bsnm{{Shibasaki}}, \binits{K.}}:
\byear{2010},
\batitle{{Types of Microwave Quasi-Periodic Pulsations in Single Flaring
  Loops}}.
\bjtitle{\solphys}
\bvolume{267},
\bfpage{329}.
\doiurl{10.1007/s11207-010-9642-0}.
\adsurl{2010SoPh..267..329K}.
\end{barticle}
\endbibitem

\bibitem[\protect\citeauthoryear{{Kuznetsov}, {Nita}, and
  {Fleishman}}{2011}]{Kuznetsov2011}
\begin{barticle}
\bauthor{\bsnm{{Kuznetsov}}, \binits{A.A.}},
\bauthor{\bsnm{{Nita}}, \binits{G.M.}},
\bauthor{\bsnm{{Fleishman}}, \binits{G.D.}}:
\byear{2011},
\batitle{{Three-dimensional Simulations of Gyrosynchrotron Emission from Mildly
  Anisotropic Nonuniform Electron Distributions in Symmetric Magnetic Loops}}.
\bjtitle{\apj}
\bvolume{742},
\bfpage{87}.
\doiurl{10.1088/0004-637X/742/2/87}.
\adsurl{2011ApJ...742...87K}.
\end{barticle}
\endbibitem

\bibitem[\protect\citeauthoryear{{Lemen} \textit{et~al.}}{2012}]{Lemen12}
\begin{barticle}
\bauthor{\bsnm{{Lemen}}, \binits{J.R.}},
\bauthor{\bsnm{{Title}}, \binits{A.M.}},
\bauthor{\bsnm{{Akin}}, \binits{D.J.}},
\bauthor{\bsnm{{Boerner}}, \binits{P.F.}},
\bauthor{\bsnm{{Chou}}, \binits{C.}},
\bauthor{\bsnm{{Drake}}, \binits{J.F.}},
\bauthor{\bsnm{{Duncan}}, \binits{D.W.}},
\bauthor{\bsnm{{Edwards}}, \binits{C.G.}},
\bauthor{\bsnm{{Friedlaender}}, \binits{F.M.}},
\bauthor{\bsnm{{Heyman}}, \binits{G.F.}},
\bauthor{\bsnm{{Hurlburt}}, \binits{N.E.}},
\bauthor{\bsnm{{Katz}}, \binits{N.L.}},
\bauthor{\bsnm{{Kushner}}, \binits{G.D.}},
\bauthor{\bsnm{{Levay}}, \binits{M.}},
\bauthor{\bsnm{{Lindgren}}, \binits{R.W.}},
\bauthor{\bsnm{{Mathur}}, \binits{D.P.}},
\bauthor{\bsnm{{McFeaters}}, \binits{E.L.}},
\bauthor{\bsnm{{Mitchell}}, \binits{S.}},
\bauthor{\bsnm{{Rehse}}, \binits{R.A.}},
\bauthor{\bsnm{{Schrijver}}, \binits{C.J.}},
\bauthor{\bsnm{{Springer}}, \binits{L.A.}},
\bauthor{\bsnm{{Stern}}, \binits{R.A.}},
\bauthor{\bsnm{{Tarbell}}, \binits{T.D.}},
\bauthor{\bsnm{{Wuelser}}, \binits{J.-P.}},
\bauthor{\bsnm{{Wolfson}}, \binits{C.J.}},
\bauthor{\bsnm{{Yanari}}, \binits{C.}},
\bauthor{\bsnm{{Bookbinder}}, \binits{J.A.}},
\bauthor{\bsnm{{Cheimets}}, \binits{P.N.}},
\bauthor{\bsnm{{Caldwell}}, \binits{D.}},
\bauthor{\bsnm{{Deluca}}, \binits{E.E.}},
\bauthor{\bsnm{{Gates}}, \binits{R.}},
\bauthor{\bsnm{{Golub}}, \binits{L.}},
\bauthor{\bsnm{{Park}}, \binits{S.}},
\bauthor{\bsnm{{Podgorski}}, \binits{W.A.}},
\bauthor{\bsnm{{Bush}}, \binits{R.I.}},
\bauthor{\bsnm{{Scherrer}}, \binits{P.H.}},
\bauthor{\bsnm{{Gummin}}, \binits{M.A.}},
\bauthor{\bsnm{{Smith}}, \binits{P.}},
\bauthor{\bsnm{{Auker}}, \binits{G.}},
\bauthor{\bsnm{{Jerram}}, \binits{P.}},
\bauthor{\bsnm{{Pool}}, \binits{P.}},
\bauthor{\bsnm{{Soufli}}, \binits{R.}},
\bauthor{\bsnm{{Windt}}, \binits{D.L.}},
\bauthor{\bsnm{{Beardsley}}, \binits{S.}},
\bauthor{\bsnm{{Clapp}}, \binits{M.}},
\bauthor{\bsnm{{Lang}}, \binits{J.}},
\bauthor{\bsnm{{Waltham}}, \binits{N.}}:
\byear{2012},
\batitle{{The Atmospheric Imaging Assembly (AIA) on the Solar Dynamics
  Observatory (SDO)}}.
\bjtitle{\solphys}
\bvolume{275},
\bfpage{17}.
\doiurl{10.1007/s11207-011-9776-8}.
\adsurl{2012SoPh..275...17L}.
\end{barticle}
\endbibitem

\bibitem[\protect\citeauthoryear{{Lin} \textit{et~al.}}{2002}]{Lin2002}
\begin{barticle}
\bauthor{\bsnm{{Lin}}, \binits{R.P.}},
\bauthor{\bsnm{{Dennis}}, \binits{B.R.}},
\bauthor{\bsnm{{Hurford}}, \binits{G.J.}},
\bauthor{\bsnm{{Smith}}, \binits{D.M.}},
\bauthor{\bsnm{{Zehnder}}, \binits{A.}},
\bauthor{\bsnm{{Harvey}}, \binits{P.R.}},
\bauthor{\bsnm{{Curtis}}, \binits{D.W.}},
\bauthor{\bsnm{{Pankow}}, \binits{D.}},
\bauthor{\bsnm{{Turin}}, \binits{P.}},
\bauthor{\bsnm{{Bester}}, \binits{M.}},
\bauthor{\bsnm{{Csillaghy}}, \binits{A.}},
\bauthor{\bsnm{{Lewis}}, \binits{M.}},
\bauthor{\bsnm{{Madden}}, \binits{N.}},
\bauthor{\bsnm{{van Beek}}, \binits{H.F.}},
\bauthor{\bsnm{{Appleby}}, \binits{M.}},
\bauthor{\bsnm{{Raudorf}}, \binits{T.}},
\bauthor{\bsnm{{McTiernan}}, \binits{J.}},
\bauthor{\bsnm{{Ramaty}}, \binits{R.}},
\bauthor{\bsnm{{Schmahl}}, \binits{E.}},
\bauthor{\bsnm{{Schwartz}}, \binits{R.}},
\bauthor{\bsnm{{Krucker}}, \binits{S.}},
\bauthor{\bsnm{{Abiad}}, \binits{R.}},
\bauthor{\bsnm{{Quinn}}, \binits{T.}},
\bauthor{\bsnm{{Berg}}, \binits{P.}},
\bauthor{\bsnm{{Hashii}}, \binits{M.}},
\bauthor{\bsnm{{Sterling}}, \binits{R.}},
\bauthor{\bsnm{{Jackson}}, \binits{R.}},
\bauthor{\bsnm{{Pratt}}, \binits{R.}},
\bauthor{\bsnm{{Campbell}}, \binits{R.D.}},
\bauthor{\bsnm{{Malone}}, \binits{D.}},
\bauthor{\bsnm{{Landis}}, \binits{D.}},
\bauthor{\bsnm{{Barrington-Leigh}}, \binits{C.P.}},
\bauthor{\bsnm{{Slassi-Sennou}}, \binits{S.}},
\bauthor{\bsnm{{Cork}}, \binits{C.}},
\bauthor{\bsnm{{Clark}}, \binits{D.}},
\bauthor{\bsnm{{Amato}}, \binits{D.}},
\bauthor{\bsnm{{Orwig}}, \binits{L.}},
\bauthor{\bsnm{{Boyle}}, \binits{R.}},
\bauthor{\bsnm{{Banks}}, \binits{I.S.}},
\bauthor{\bsnm{{Shirey}}, \binits{K.}},
\bauthor{\bsnm{{Tolbert}}, \binits{A.K.}},
\bauthor{\bsnm{{Zarro}}, \binits{D.}},
\bauthor{\bsnm{{Snow}}, \binits{F.}},
\bauthor{\bsnm{{Thomsen}}, \binits{K.}},
\bauthor{\bsnm{{Henneck}}, \binits{R.}},
\bauthor{\bsnm{{McHedlishvili}}, \binits{A.}},
\bauthor{\bsnm{{Ming}}, \binits{P.}},
\bauthor{\bsnm{{Fivian}}, \binits{M.}},
\bauthor{\bsnm{{Jordan}}, \binits{J.}},
\bauthor{\bsnm{{Wanner}}, \binits{R.}},
\bauthor{\bsnm{{Crubb}}, \binits{J.}},
\bauthor{\bsnm{{Preble}}, \binits{J.}},
\bauthor{\bsnm{{Matranga}}, \binits{M.}},
\bauthor{\bsnm{{Benz}}, \binits{A.}},
\bauthor{\bsnm{{Hudson}}, \binits{H.}},
\bauthor{\bsnm{{Canfield}}, \binits{R.C.}},
\bauthor{\bsnm{{Holman}}, \binits{G.D.}},
\bauthor{\bsnm{{Crannell}}, \binits{C.}},
\bauthor{\bsnm{{Kosugi}}, \binits{T.}},
\bauthor{\bsnm{{Emslie}}, \binits{A.G.}},
\bauthor{\bsnm{{Vilmer}}, \binits{N.}},
\bauthor{\bsnm{{Brown}}, \binits{J.C.}},
\bauthor{\bsnm{{Johns-Krull}}, \binits{C.}},
\bauthor{\bsnm{{Aschwanden}}, \binits{M.}},
\bauthor{\bsnm{{Metcalf}}, \binits{T.}},
\bauthor{\bsnm{{Conway}}, \binits{A.}}:
\byear{2002},
\batitle{{The Reuven Ramaty High-Energy Solar Spectroscopic Imager (RHESSI)}}.
\bjtitle{\solphys}
\bvolume{210},
\bfpage{3}.
\doiurl{10.1023/A:1022428818870}.
\adsurl{2002SoPh..210....3L}.
\end{barticle}
\endbibitem

\bibitem[\protect\citeauthoryear{{Magara} \textit{et~al.}}{1996}]{Magara1996}
\begin{barticle}
\bauthor{\bsnm{{Magara}}, \binits{T.}},
\bauthor{\bsnm{{Mineshige}}, \binits{S.}},
\bauthor{\bsnm{{Yokoyama}}, \binits{T.}},
\bauthor{\bsnm{{Shibata}}, \binits{K.}}:
\byear{1996},
\batitle{{Numerical Simulation of Magnetic Reconnection in Eruptive Flares}}.
\bjtitle{\apj}
\bvolume{466},
\bfpage{1054}.
\doiurl{10.1086/177575}.
\adsurl{1996ApJ...466.1054M}.
\end{barticle}
\endbibitem

\bibitem[\protect\citeauthoryear{{Morgachev}, {Kuznetsov}, and
  {Melnikov}}{2014}]{Morgachev2014}
\begin{barticle}
\bauthor{\bsnm{{Morgachev}}, \binits{A.S.}},
\bauthor{\bsnm{{Kuznetsov}}, \binits{S.A.}},
\bauthor{\bsnm{{Melnikov}}, \binits{V.F.}}:
\byear{2014},
\batitle{{Radio diagnostics of the solar flaring loop parameters by the forward
  fitting method}}.
\bjtitle{Geomagnetism and Aeronomy}
\bvolume{54},
\bfpage{933}.
\doiurl{10.1134/S0016793214070081}.
\adsurl{2014Ge\%26Ae..54..933M}.
\end{barticle}
\endbibitem

\bibitem[\protect\citeauthoryear{{Nakajima} \textit{et~al.}}{1994}]{Nakajima94}
\begin{barticle}
\bauthor{\bsnm{{Nakajima}}, \binits{H.}},
\bauthor{\bsnm{{Nishio}}, \binits{M.}},
\bauthor{\bsnm{{Enome}}, \binits{S.}},
\bauthor{\bsnm{{Shibasaki}}, \binits{K.}},
\bauthor{\bsnm{{Takano}}, \binits{T.}},
\bauthor{\bsnm{{Hanaoka}}, \binits{Y.}},
\bauthor{\bsnm{{Torii}}, \binits{C.}},
\bauthor{\bsnm{{Sekiguchi}}, \binits{H.}},
\bauthor{\bsnm{{Bushimata}}, \binits{T.}},
\bauthor{\bsnm{{Kawashima}}, \binits{S.}},
\bauthor{\bsnm{{Shinohara}}, \binits{N.}},
\bauthor{\bsnm{{Irimajiri}}, \binits{Y.}},
\bauthor{\bsnm{{Koshiishi}}, \binits{H.}},
\bauthor{\bsnm{{Kosugi}}, \binits{T.}},
\bauthor{\bsnm{{Shiomi}}, \binits{Y.}},
\bauthor{\bsnm{{Sawa}}, \binits{M.}},
\bauthor{\bsnm{{Kai}}, \binits{K.}}:
\byear{1994},
\batitle{{The Nobeyama radioheliograph.}}
\bjtitle{IEEE Proceedings}
\bvolume{82},
\bfpage{705}.
\adsurl{1994IEEEP..82..705N}.
\end{barticle}
\endbibitem

\bibitem[\protect\citeauthoryear{{Nita} \textit{et~al.}}{2015}]{Nita2015}
\begin{barticle}
\bauthor{\bsnm{{Nita}}, \binits{G.M.}},
\bauthor{\bsnm{{Fleishman}}, \binits{G.D.}},
\bauthor{\bsnm{{Kuznetsov}}, \binits{A.A.}},
\bauthor{\bsnm{{Kontar}}, \binits{E.P.}},
\bauthor{\bsnm{{Gary}}, \binits{D.E.}}:
\byear{2015},
\batitle{{Three-dimensional Radio and X-Ray Modeling and Data Analysis
  Software: Revealing Flare Complexity}}.
\bjtitle{\apj}
\bvolume{799},
\bfpage{236}.
\doiurl{10.1088/0004-637X/799/2/236}.
\adsurl{2015ApJ...799..236N}.
\end{barticle}
\endbibitem

\bibitem[\protect\citeauthoryear{{Pinto} \textit{et~al.}}{2016}]{Pinto2016}
\begin{barticle}
\bauthor{\bsnm{{Pinto}}, \binits{R.F.}},
\bauthor{\bsnm{{Gordovskyy}}, \binits{M.}},
\bauthor{\bsnm{{Browning}}, \binits{P.K.}},
\bauthor{\bsnm{{Vilmer}}, \binits{N.}}:
\byear{2016},
\batitle{{Thermal and non-thermal emission from reconnecting twisted coronal
  loops}}.
\bjtitle{\aap}
\bvolume{585},
\bfpage{A159}.
\doiurl{10.1051/0004-6361/201526633}.
\adsurl{2016A\%26A...585A.159P}.
\end{barticle}
\endbibitem

\bibitem[\protect\citeauthoryear{{Rudenko} and {Myshyakov}}{2009}]{Rudenko2009}
\begin{barticle}
\bauthor{\bsnm{{Rudenko}}, \binits{G.V.}},
\bauthor{\bsnm{{Myshyakov}}, \binits{I.I.}}:
\byear{2009},
\batitle{{Analysis of Reconstruction Methods for Nonlinear Force-Free Fields}}.
\bjtitle{\solphys}
\bvolume{257},
\bfpage{287}.
\doiurl{10.1007/s11207-009-9389-7}.
\adsurl{2009SoPh..257..287R}.
\end{barticle}
\endbibitem

\bibitem[\protect\citeauthoryear{{Scherrer}
  \textit{et~al.}}{2012}]{Scherrer2012}
\begin{barticle}
\bauthor{\bsnm{{Scherrer}}, \binits{P.H.}},
\bauthor{\bsnm{{Schou}}, \binits{J.}},
\bauthor{\bsnm{{Bush}}, \binits{R.I.}},
\bauthor{\bsnm{{Kosovichev}}, \binits{A.G.}},
\bauthor{\bsnm{{Bogart}}, \binits{R.S.}},
\bauthor{\bsnm{{Hoeksema}}, \binits{J.T.}},
\bauthor{\bsnm{{Liu}}, \binits{Y.}},
\bauthor{\bsnm{{Duvall}}, \binits{T.L.}},
\bauthor{\bsnm{{Zhao}}, \binits{J.}},
\bauthor{\bsnm{{Title}}, \binits{A.M.}},
\bauthor{\bsnm{{Schrijver}}, \binits{C.J.}},
\bauthor{\bsnm{{Tarbell}}, \binits{T.D.}},
\bauthor{\bsnm{{Tomczyk}}, \binits{S.}}:
\byear{2012},
\batitle{{The Helioseismic and Magnetic Imager (HMI) Investigation for the
  Solar Dynamics Observatory (SDO)}}.
\bjtitle{\solphys}
\bvolume{275},
\bfpage{207}.
\doiurl{10.1007/s11207-011-9834-2}.
\adsurl{2012SoPh..275..207S}.
\end{barticle}
\endbibitem

\bibitem[\protect\citeauthoryear{{Sharykin} and
  {Kuznetsov}}{2016}]{Sharykin2016}
\begin{barticle}
\bauthor{\bsnm{{Sharykin}}, \binits{I.N.}},
\bauthor{\bsnm{{Kuznetsov}}, \binits{A.A.}}:
\byear{2016},
\batitle{{Modelling of Nonthermal Microwave Emission from Twisted Magnetic
  Loops}}.
\bjtitle{\solphys}
\bvolume{291},
\bfpage{1341}.
\doiurl{10.1007/s11207-016-0917-y}.
\adsurl{2016SoPh..291.1341S}.
\end{barticle}
\endbibitem

\bibitem[\protect\citeauthoryear{{Syrovatskii} and
  {Shmeleva}}{1972}]{Syrovatskii1972}
\begin{barticle}
\bauthor{\bsnm{{Syrovatskii}}, \binits{S.I.}},
\bauthor{\bsnm{{Shmeleva}}, \binits{O.P.}}:
\byear{1972},
\batitle{{Heating of Plasma by High-Energy Electrons, and Nonthermal X-Ray
  Emission in Solar Flares}}.
\bjtitle{\sovast}
\bvolume{16},
\bfpage{273}.
\adsurl{1972SvA....16..273S}.
\end{barticle}
\endbibitem

\bibitem[\protect\citeauthoryear{{Titov} and {D{\'e}moulin}}{1999}]{Titov1999}
\begin{barticle}
\bauthor{\bsnm{{Titov}}, \binits{V.S.}},
\bauthor{\bsnm{{D{\'e}moulin}}, \binits{P.}}:
\byear{1999},
\batitle{{Basic topology of twisted magnetic configurations in solar flares}}.
\bjtitle{\aap}
\bvolume{351},
\bfpage{707}.
\adsurl{1999A\%26A...351..707T}.
\end{barticle}
\endbibitem

\bibitem[\protect\citeauthoryear{{Tsuneta}}{1997}]{Tsuneta1997}
\begin{barticle}
\bauthor{\bsnm{{Tsuneta}}, \binits{S.}}:
\byear{1997},
\batitle{{Moving Plasmoid and Formation of the Neutral Sheet in a Solar
  Flare}}.
\bjtitle{\apj}
\bvolume{483},
\bfpage{507}.
\adsurl{1997ApJ...483..507T}.
\end{barticle}
\endbibitem

\bibitem[\protect\citeauthoryear{{Wheatland}, {Sturrock}, and
  {Roumeliotis}}{2000}]{Wheatland2000}
\begin{barticle}
\bauthor{\bsnm{{Wheatland}}, \binits{M.S.}},
\bauthor{\bsnm{{Sturrock}}, \binits{P.A.}},
\bauthor{\bsnm{{Roumeliotis}}, \binits{G.}}:
\byear{2000},
\batitle{{An Optimization Approach to Reconstructing Force-free Fields}}.
\bjtitle{\apj}
\bvolume{540},
\bfpage{1150}.
\doiurl{10.1086/309355}.
\adsurl{2000ApJ...540.1150W}.
\end{barticle}
\endbibitem

\bibitem[\protect\citeauthoryear{{Zheleznyakov} and
  {Zlotnik}}{1964}]{Zheleznyakov1964}
\begin{barticle}
\bauthor{\bsnm{{Zheleznyakov}}, \binits{V.V.}},
\bauthor{\bsnm{{Zlotnik}}, \binits{E.Y.}}:
\byear{1964},
\batitle{{Polarization of Radio Waves Passing through a Transverse Magnetic
  Field Region in the Solar Corona}}.
\bjtitle{\sovast}
\bvolume{7},
\bfpage{485}.
\adsurl{1964SvA.....7..485Z}.
\end{barticle}
\endbibitem

\end{thebibliography}

\end{article}
\end{document}